\documentclass[%
reprint,
superscriptaddress,
showpacs,
amsmath,
amssymb,
aps,
prl,
nobalancelastpage]{revtex4-2}

\usepackage{color}
\usepackage{graphicx}
\usepackage[utf8]{inputenc}
\usepackage[english]{babel} 
\usepackage{microtype}
\usepackage{amsmath}
\usepackage{amssymb}
\usepackage{bm}
\usepackage{bbm}
\usepackage{dsfont}
\usepackage{eucal}
\usepackage[colorlinks,allcolors=black]{hyperref}
\usepackage[capitalise]{cleveref}

\graphicspath{{img/}}

\begin{document}

\title{Reinforcement Learning with thermal fluctuations
at the nano-scale}

\author{Francesco Boccardo}
\affiliation{Institut Lumi\`ere Mati\`ere, UMR5306, Universit\'e Lyon 1 - CNRS, Villeurbanne, France}
\affiliation{MaLGa, Department of Civil, Chemical and Environmental Engineering, University of Genoa, Genoa, Italy}
\author{Olivier Pierre-Louis}
\email{olivier.pierre-louis@univ-lyon1.fr}
\affiliation{Institut Lumi\`ere Mati\`ere, UMR5306, Universit\'e Lyon 1 - CNRS, Villeurbanne, France}
\date{\today}

\begin{abstract}
Reinforcement Learning 
offers a framework to learn 
to choose actions in order to achieve 
some goal.
However, at the nano-scale, thermal fluctuations hamper the learning process. 
We analyze this regime using the general framework of  Markov Decision Processes, 
which applies to a wide variety of problems from nano-navigation 
to nano-machine actuation. We show 
that at the nan-oscale, while optimal actions should bring an improvement proportional to the small ratio of the applied force times a length-scale over the temperature, the learned improvement is smaller and proportional to the square of this small ratio.
Consequently, the efficiency of learning, which compares the learning improvement to the theoretical optimal improvement, drops to zero.
Nevertheless, these limitations can be circumvented by using actions learned at a lower temperature.
These results are illustrated with simulations of the control 
of the shape of small particle clusters.
\end{abstract}

\maketitle

Reinforcement Learning (RL),
the process by which an agent learns to choose actions on a dynamical system to
maximize rewards from this system,
is one of the main
Machine Learning paradigms~\cite{Sutton1998}.
Following its long-established use for games~\cite{Mnih2013} and robotics~\cite{Kober2013,Zhao2020},
RL has recently paved the way for
many breakthroughs in the control of systems that are constrained 
by various physical and biological environments. A few examples include the control of 
fluid flows~\cite{Garnier2021,Beintema2020},
biological and artificial navigation~\cite{Singh2023,Nasiri2022,Cichos2020,Yang2020,Novati2019,Schneider2019,Reddy2018,Reddy2016,Colabrese2017},
organization of active assemblies~\cite{Durve2020,MuinosLandin2021},
and order and shape of colloidal clusters~\cite{Zhang2020}.
Many of these systems 
involve objects with very small sizes.
However, the development of RL for micro- and nano-meter scale objects
faces challenging regimes where thermal fluctuations lead to erratic Brownian motion that dominates the dynamics, 
and actions then only produce a small bias in stochastic rewards.
For instance, nano-motors must be actuated
in a robust way within these fluctuating environments~\cite{Fletcher2005,Mallouk2009,Shao2018,Robertson2018}.
In addition, external fields
to manipulate colloids and nano-machines must be weak
to achieve nano-robot navigation for drug delivery 
or remote-controlled surgery~\cite{Zhang2022,Li2016,Li2018,Zarei2018} 
without damaging the surrounding soft living environment.
These systems and the associated emerging technologies 
call for a better fundamental understanding 
of the performance of RL 
when the actuation forces are small as compared to thermal fluctuations.

While a large body of work has been devoted
to the robustness of learning in noisy environments~\cite{Nilim2005,Iyengar2005,Wang2020,Brunke2022,Dridi2015},
only a few have focused on the role of thermal fluctuations that are
 ubiquitous at the nano-scale.
Previous works on navigation in gridworlds
(where space is represented by a finite lattice) suggest that learning is more 
difficult when increasing temperature~\cite{MuinosLandin2021,Larchenko2021}.
Temperature indeed has a strong effect on the speed of the dynamics,
which is usually faster at high temperatures and slower at low temperatures.
This leads to difficulties in observation, as transitions become too fast
or too slow to be observed in experiments.
In the following, we show that beyond these difficulties in observability,
there are intrinsic physical constraints that make the efficiency of RL vanish 
in the presence of strong thermal fluctuations.

We investigate this question using simulations of a specific problem that can be considered
as a prototype of a nano-robot, where the shape 
of a fluctuating few-particle cluster is controlled with a macroscopic field~\cite{Boccardo2022}.
The macroscopic field biases the stochastic 
configurational changes of the cluster,
as expected for atomic or colloidal clusters 
in the presence, e.g., of an electric field~\cite{Zhang2020,Kuhn2005,Mahadevan1999,PierreLouis2000,Curiotto2019},
or light~\cite{McCormack2018}.
Our aim is to set the macroscopic field
as a function of the observed shape 
to reach  an arbitrary target shape
in minimum time. 
The experimental realization of this system
when downscaling particle sizes towards the atomic size is a challenge
due to the weakness of available driving forces such as electromigration~\cite{Boccardo2022}.
This problem can be formulated within the general framework of Markov Decision Processes,
i.e., a Markov chain where the choice of an action---called the policy---is made as a function of the observed state. 
Hence, our results 
can be transposed to other RL problems
and pertain to all systems that can be 
modeled with Markov Decision Processes~\cite{Sutton1998,MuinosLandin2021}, 
including 
for example actuation of 
molecular machines, navigation, and controlled assembly.

In order to grasp the effect of temperature on learning,
we do not employ the most sophisticated and powerful RL methods.
Instead, we use two of the most elementary ones~\cite{Sutton1998}: 
Monte Carlo Learning (MCL) and Q-Learning (QL),
which are based on $\varepsilon$-greedy policies.
These are standard methods to deal with discrete sets of states and actions,
which is our focus in the following.
We find that the amount of control achieved by RL depends 
crucially on the dimensionless ratio between
the work of the applied force during an elementary move 
and the thermal energy. 
When this ratio is small, corresponding to small forces, small scales, 
or high temperatures, 
the efficiency of RL is proportional to it. 
However, this inefficiency of learning at high temperatures
can be circumvented by
using actions learned at a lower temperature.

Our analysis is based on
the comparison of RL to the optimal solution of the control problem.
This solution is obtained with Dynamic 
Programming (DP)~\cite{Boccardo2022,Xue2014,Xue2013,Banerjee2009,Sutton1998},
a model-based method 
that relies on the full knowledge of the laws that govern the system,
while RL is a model-free approach that only relies on the observation of the response of the system to some actions without prior knowledge on the governing laws.

%%%%%%%%%%%%%%%%%%%%%%%%%%%%%%%%%%%%%%%%%%%%%%%%%%%%%%%%%%%%
\paragraph{Cluster model.}
%%%%%%%%%%%%%%%%%%%%%%%%%%%%%%%%%%%%%%%%%%%%%%%%%%%%%%%%%%%%

The dynamics of a fluctuating few-particle cluster are modeled 
using on-lattice edge-diffusion dynamics~\cite{Boccardo2022,PierreLouis2000,Glasstone1941,Liu1998}.
Edge diffusion was observed in metal atomic monolayer islands~\cite{Giesen2001,Tao2010}, 
and colloids~\cite{Hubartt2015}.
Our approach can be readily applied to other
types of dynamics for clusters or molecules that preserve the number
of particles, such as dislocation-induced
events in metal clusters, colloidal clusters and Wigner crystals,
detachment-diffusion-reattachment dynamics
inside vacancies in particle monolayers~\cite{Boccardo2022b,PierreLouis2000}, 
or dynamics of polymer and proteins~\cite{Swope2004}.

In this model, particles hop to nearest neighbor sites 
along the cluster edge and cannot detach from the cluster or break the cluster~\cite{Boccardo2022}. 
The particle hopping rate follows an Arrhenius law~\cite{Boccardo2022,Glasstone1941,Liu1998}
\begin{align}\label{eq:TST_hopping}
    \gamma = \nu \exp[-(nJ - \mathbf{F}\cdot\bm{u}d)/k_BT],
\end{align}
where $J$ is the bond energy, $n$
is the number of bonds of the particle before hopping,
$k_BT$ is the thermal energy, $\mathbf{F}$ is the macroscopic force field,
$d$ is the lattice constant,
and $\bm{u}d$ is the vector
from the initial to the saddle position of the moving particle.
For the sake of simplicity, we assume that the saddle
point of the diffusion energy landscape is half-way between
the initial and final positions~\cite{Boccardo2022}.

%%%%%%%%%%%%%%%%%%%%%%%%%%%%%%%%%%%%%%%%%%%%%%%%%%%%%%%%%%%%
\paragraph{RL algorithms.}
%%%%%%%%%%%%%%%%%%%%%%%%%%%%%%%%%%%%%%%%%%%%%%%%%%%%%%%%%%%%

In the language of 
Markov Decision Processes~\cite{Sutton1998}, the configuration or shape of the cluster
is the state $s$ of the system.
Moreover, we consider a discrete set of actions labelled
by the index $a$.
For simplicity, we consider only 3 possible actions $a=-1,0,+1$, 
that respectively correspond to setting the force to the left,
to zero, or to the right, i.e. $\mathbf{F}=a F {\mathbf e}_x$,
where $F>0$ and ${\mathbf e}_x$ is the unit vector
along the $(10)$ lattice direction.

The policy $\pi(a|s)$
is the probability to choose action $a$ in state $s$,
and the reward is minus the residence time 
$\hat{r}=-\hat{t}$ in this state.
Here and in the following, a hat indicates a stochastic variable.
An episode is a sample of the dynamics
consisting of a list of states, actions, and rewards
$\{ \hat{s}_{k-1},\hat{a}_{k-1},\hat{r}_k; k=1,\dots,K \}$.
The sum of all future rewards is 
$\hat{g}_k=\sum_{p=k+1}^K\hat{r}_{p}$. 
Episodes terminate when the state $s$ reaches the target state $\bar{s}$ or when 
$k$ reaches the maximum allowed number of steps $M$.
Assuming that $M$ is large enough
for the target state  $\bar{s}$ to be reached with high probability
before the end of the episode,
minus the average of $\hat{g}$ is a good approximation of the
expected first passage time $\tau_\pi(s;\bar{s})$,
i.e.
\begin{align}
    \tau_{\pi}(s;\bar{s})\approx-\mathbb{E}_{\pi}[\hat{g}_k|\hat{s}_k=s],
    \label{eq:expected_first_passage_time}
\end{align}
where $\mathbb{E}_\pi[\cdot]$ is the expected value under the policy $\pi$.

The optimal policy $\pi_*$
is a deterministic policy which minimizes the 
first passage time to the target state: $\pi_*\in\mathrm{argmin}_\pi\tau_\pi(s;\bar{s})$.
The optimal first passage times $\tau_*(s;\bar{s})$ obey
the Bellman optimality equation~\cite{Bellman1957,Sutton1998}
\begin{align}\label{eq:Bellman_optim}
  \tau_*(s;\bar{s})=\min_a\Bigl[t(s,a)+\sum_{s'\in{\cal B}_s}p(s'|s,a)\tau_*(s;\bar{s})\Bigr],
\end{align}
where {${\cal B}_s$ is the set of states that can be reached from
$s$ in one move,} $t(s,a)$ and $p(s'|s,a)$ are respectively 
the average residence time 
and the transition probability to state $s'$
when the system is in state $s$
with the action $a$.
This equation can be solved numerically by DP~\cite{Sutton1998,Boccardo2022},
which here consists in iterating~\cref{eq:Bellman_optim}.
Optimal policies for 5 and 7-particle clusters
are shown in { Supplemental Material (SM) Fig.~S1 and S2}.

\begin{figure}[ht]
    \centering
    \includegraphics[width=0.8\columnwidth]{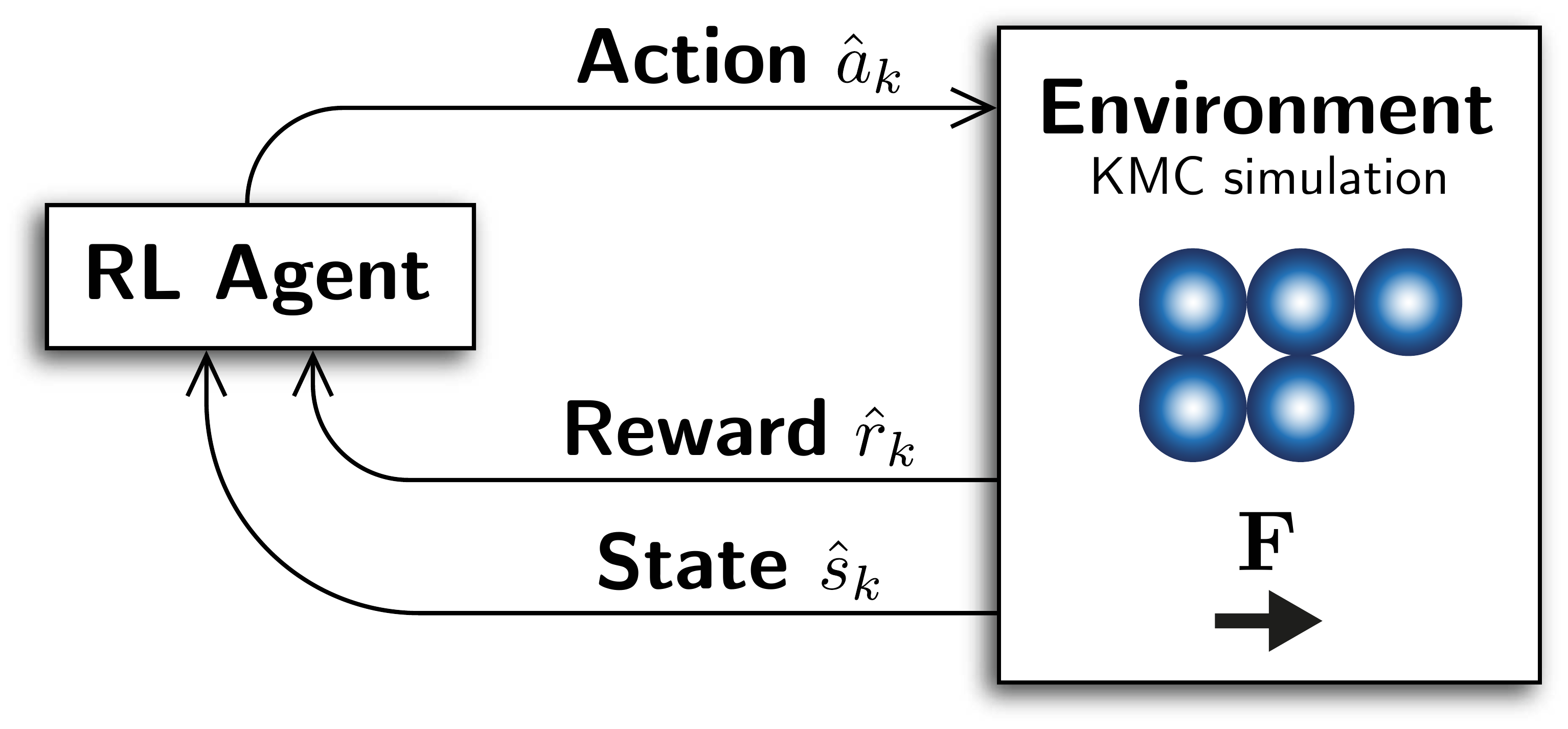}
    \caption{Schematic of the agent-environment interface in RL.}
    \label{fig:1}
\end{figure}

In contrast, RL algorithms 
aim to find $\tau_*(s;\bar{s})$ and $\pi_*$
from a set of episodes where they choose the actions without prior knowledge of the model,
i.e., of $t(s,a)$ and $p(s'|s,a)$. 
Here, RL only observes trajectories produced by Kinetic Monte Carlo (KMC), as schematized in \cref{fig:1}.
We use two well-known RL algorithms~\cite{Sutton1998}
based on the evaluation of the 
action-value function $q_\pi(s,a;\bar{s})$,
which is the expected value of $\hat{g}$ starting 
from state $s$, taking the action $a$ first, 
and then following the policy $\pi$ for subsequent actions.
The first one, MCL, evaluates $q_\pi(s,a;\bar{s})$ 
from a direct estimate of \cref{eq:expected_first_passage_time} 
using averages over the trajectories that lead to the 
target in the episodes. In contrast, QL updates  an approximation of $q_\pi(s,a;\bar{s})$
so as to reduce the so-called Time-Difference error~\cite{Sutton1998}.

In both MCL and QL, the actions are chosen
in a such a way to explore the states
using an $\varepsilon$-greedy policy: 
as the exploration parameter $\varepsilon$ decreases during learning,
actions are chosen very randomly initially,
but gradually more greedily,
i.e., in a way that 
minimizes the evaluation
of the first passage time to target based on the
current approximate estimate of the action-value function.
This gradual decrease of the randomness of the policy
reflects the standard Machine Learning trade-off between exploration and exploitation.
At the end of the simulation, 
we obtain a deterministic
policy that approximates an optimal one.
Details 
of these well-known algorithms are provided in 
{ SM Sec. II}. 

\begin{figure*}[ht]
    \centering
    \includegraphics[width=0.33\textwidth]{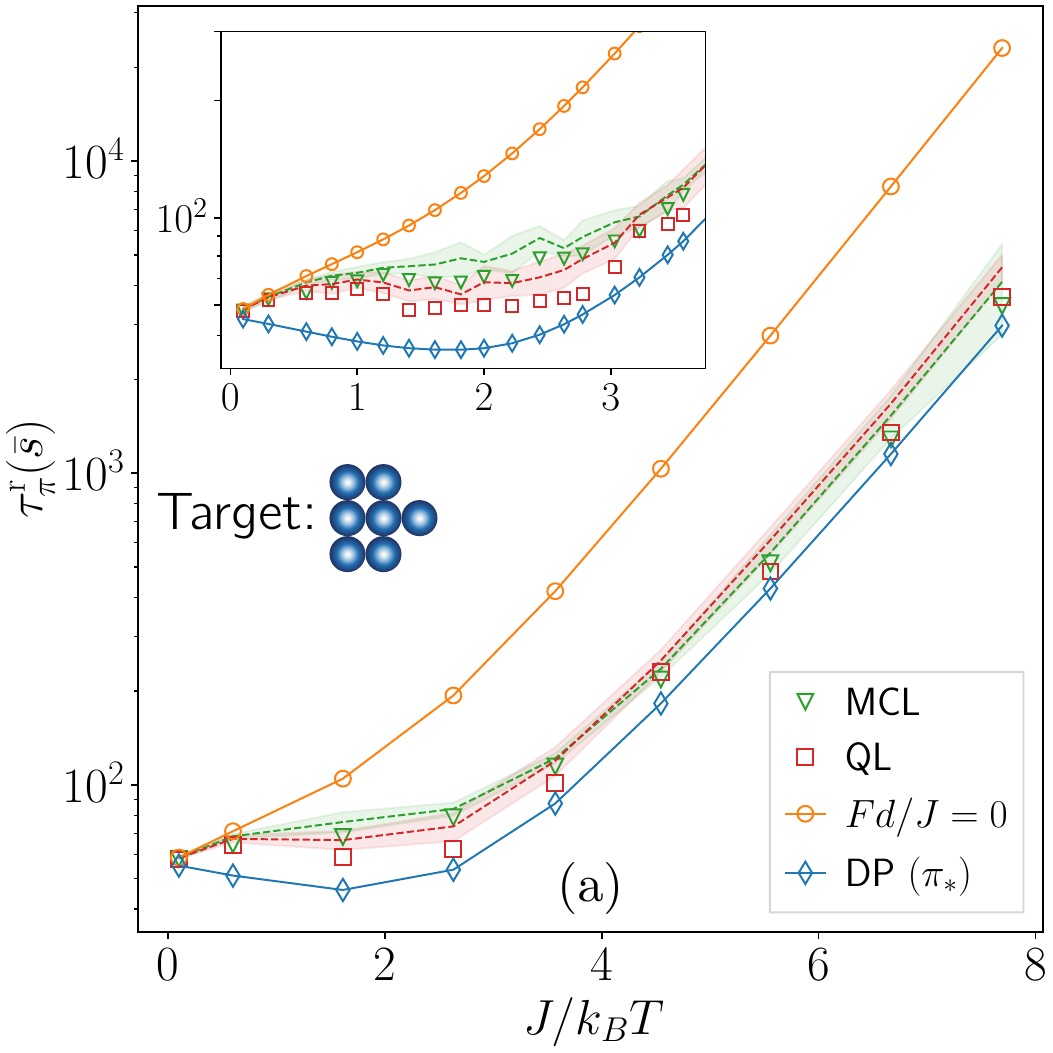}\hfill
    \includegraphics[width=0.33\textwidth]{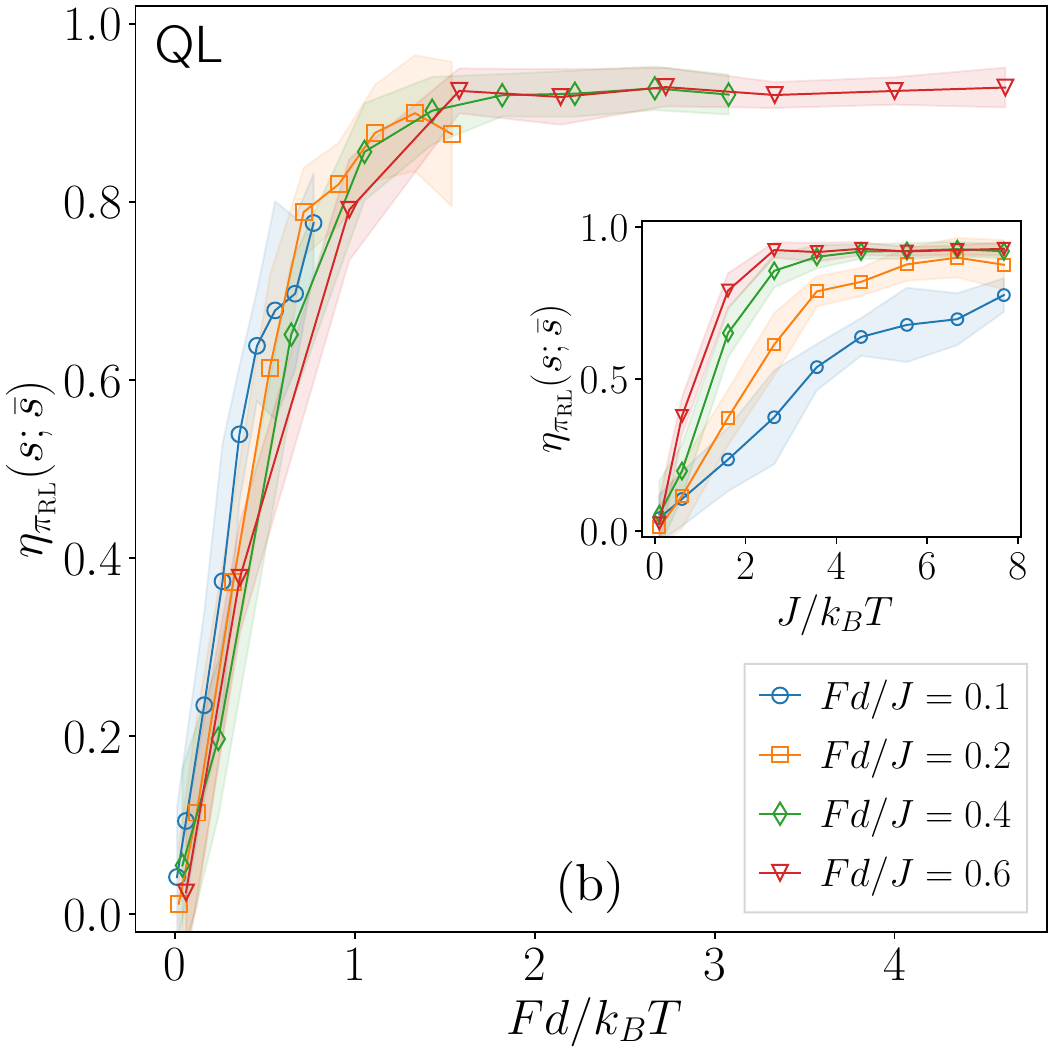}\hfill
    \includegraphics[width=0.33\textwidth]{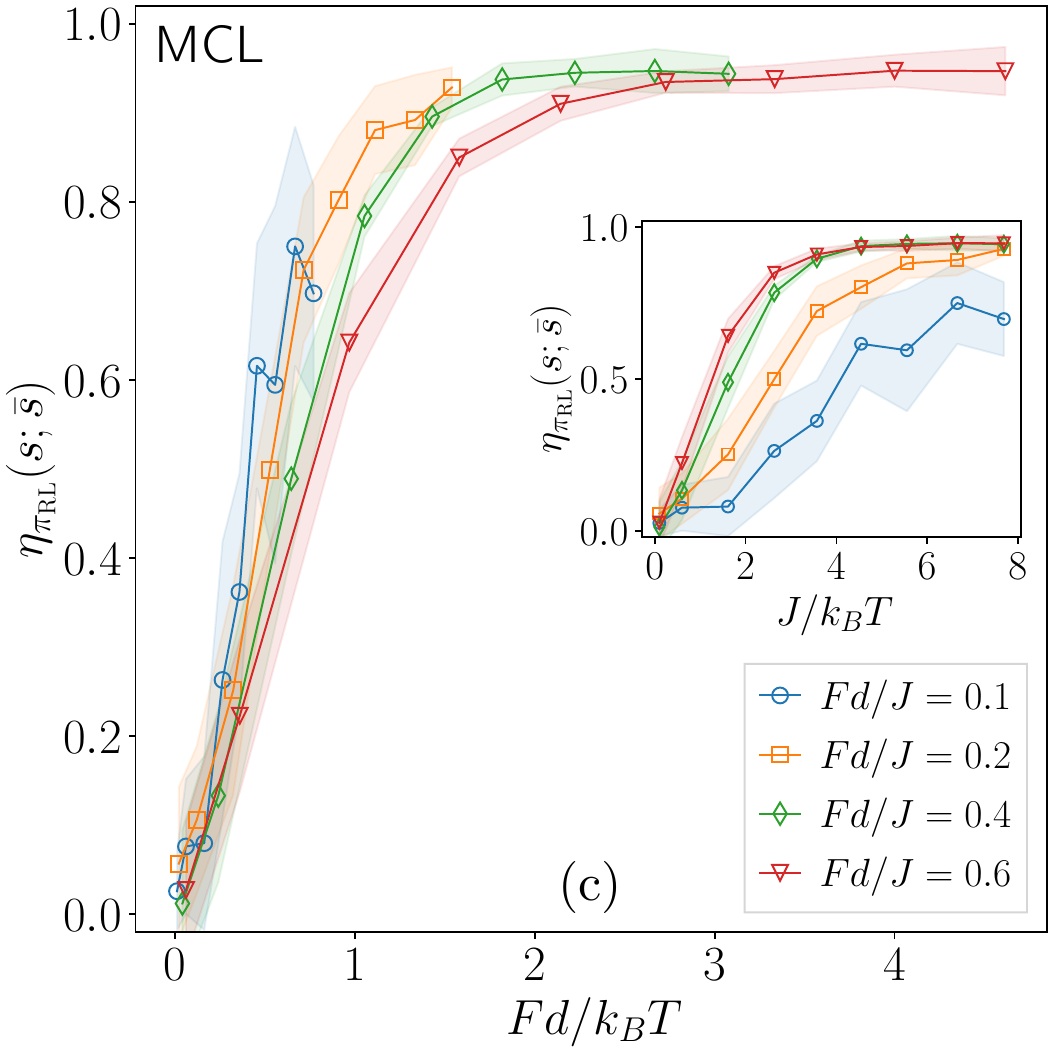}
    \caption{Learning efficiency.
    (a) Return time to target for: no force, DP, MCL and QL policies (evaluated with DP) as a function of $J/k_BT$, with fixed $Fd/k_BT=0.4$. Inset: zoom in the intermediate/high temp regime. The symbols represent the shortest return time out of 10 independent learning runs and the dashed line corresponds to the average.
    (b) Average learning efficiency $\eta_{\pi_\text{RL}}(\bar{s})$ 
    (defined in \cref{eq:learning_efficiency}) as a function of
    $Fd/k_BT$ for QL. Inset: as a function of $J/k_BT$.
    (c) Learning efficiency $\eta_{\pi_\text{RL}}(\bar{s})$ as a function of $Fd/k_BT$ for MCL. Inset: as a function of $J/k_BT$. 
    In (a,b,c) the shaded area represents the standard deviation over the 10 learning runs.
    }
    \label{fig:2}
\end{figure*}

Following Refs.~\cite{Boccardo2022,Boccardo2022b,Boccardo2022c},
we use the return time to target $\tau^{\mathrm r}(\bar{s})$,
defined as the time of first return to the target
after leaving it,
as a simple representative value of the first passage times
$\tau_\pi(s,\bar{s})$
from all the states $s$.
We use the improvement of $\tau^{\mathrm r}(\bar{s})$
to define our learning convergence criterion. To avoid limitations due to
observability, this criterion is unaware of elapsed physical time,
and is only based on the number of episodes,
as discussed in { SM Sec. II}. 
Furthermore, to keep simulation costs down, we focus on clusters with at most ten particles~\cite{Boccardo2022}.

%%%%%%%%%%%%%%%%%%%%%%%%%%%%%%%%%%%%%%%%%%%%%%%%%%%%%%%%%%%%
\paragraph{Vanishing learning at high temperatures and small forces.}
%%%%%%%%%%%%%%%%%%%%%%%%%%%%%%%%%%%%%%%%%%%%%%%%%%%%%%%%%%%%

In \cref{fig:2}~(a), the 
return time to target obtained with DP, MCL and QL
is shown for a given target state as a function 
of the inverse temperature $J/k_BT$
for a fixed value of \mbox{$Fd/J=0.4$}. 
The case of zero force is also shown, and will
serve as a reference for the value of
the return time without any optimization (a random policy
could also be used, as discussed in { SM Sec. IV}). 
Strikingly, while the return time obtained with MCL and QL at low temperatures is 
close to the optimal result
provided by DP, no learning is obtained at high temperatures.

The effectiveness of learning can be 
quantified by the efficiency of the learned policy $\pi_\mathrm{RL}$,
defined as the ratio of the reduction of the return time
obtained with $\pi_\mathrm{RL}$ over the optimal value of this reduction
\begin{equation}\label{eq:learning_efficiency}
    \eta_{\pi_\mathrm{RL}}(\bar{s}) = 
    \dfrac{ \tau^{\mathrm r}_{\mathrm{0}}({\bar{s}}) - \tau^{\mathrm r}_{\pi_\mathrm{RL}}({\bar{s}}) }
    { \tau^{\mathrm r}_{\mathrm{0}}({\bar{s}}) - \tau^{\mathrm r}_{*}({\bar{s}}) }\,,
\end{equation}
where $\tau^{\mathrm r}_{\mathrm{0}}({\bar{s}})$,
$\tau^{\mathrm r}_{*}({\bar{s}})$
and $\tau^{\mathrm r}_{\pi_\mathrm{RL}}({\bar{s}})$
are the return times with zero force, optimal policy obtained by DP, 
and RL policy respectively.
In \cref{fig:2}~(b) and (c), all efficiencies for QL and MCL for different values of $Fd/J$ are seen to drop to zero when $Fd/k_BT$ is small.
This drop of efficiency is also seen in the return time for
other targets and for all first passage times from 
other states, as shown in { SM Fig.~S3 and S4}.

A heuristic reasoning 
can rationalize this drop.
The effect of the actions
is generically expected to be included via the work
of the forces $\mathbf{F}\cdot \mathbf{u} d\sim Fda$ associated to 
a displacement $\mathbf{u} d$, as in \cref{eq:TST_hopping}.
Indeed, an expansion of the rates~\cref{eq:TST_hopping} 
to linear order in the small parameter $Fd/k_BT$
leads to a first-order correction proportional to $Fda/k_BT$.
Hence, for a given change in the policy that changes the actions $a$ 
in the states, the change of the residence times 
 are proportional to $Fd/k_BT$.
Therefore, the optimal reduction of the first passage
times---which are a sum of future residence times---is also proportional to $Fd/k_BT$, i.e. 
$\tau^{\mathrm r}_{\mathrm{0}}({\bar{s}}) - \tau^{\mathrm r}_{*}({\bar{s}})\sim O(Fd/k_BT)$.
However, during learning, the policy itself
is obtained from an estimate of the action-value function $q_\text{RL}$
which is dominated by the noisiness of the learning process. 
For small $Fd/k_BT$, the probability to find the optimal action
from the noisy RL estimate of $q_\text{RL}$
is again proportional to $Fd/k_BT$. Therefore, the expected 
value of the actions $a$ is proportional to $Fd/k_BT$,
and the expected value of the change in the rewards is proportional to 
$Fda/k_BT\sim (Fd/k_BT)^2$. We then have
$ \tau^{\mathrm r}_{\mathrm{0}}({\bar{s}}) - \tau^{\mathrm r}_{\mathrm{RL}}({\bar{s}})\sim O(Fd/k_BT)^2$.
As a consequence, the efficiency $\eta_{\pi_\mathrm{RL}}(\bar{s})$ drops and is proportional
to $Fd/k_BT$.  To confirm this heuristic analysis, a rigorous
derivation of the drop of efficiency based on 
an expansion of the return time to first order in the small parameter $Fd/k_BT$
is reported in { SM Sec. IV}.

Our results show that 
the efficiency drops at small $Fd/k_BT$. Since in practice this means 
that learning does not converge, such a statement seems to be in contradiction
with well known-proofs of convergence 
for the RL methods that we use~\cite{Sutton1998}. 
However, there is no contradiction because it is always possible to
use a stronger convergence criterion to improve learning. 
In other words, the resources needed to converge at small $Fd/k_BT$ 
increase, and in practical applications the difficulty
of learning in this regime therefore has to be faced.

%%%%%%%%%%%%%%%%%%%%%%%%%%%%%%%%%%%%%%%%%%%%%%%%%%%%%%%%%%%%
\paragraph{Transfer learning from low temperatures.}
%%%%%%%%%%%%%%%%%%%%%%%%%%%%%%%%%%%%%%%%%%%%%%%%%%%%%%%%%%%%

\begin{figure}[ht]
    \centering  
    \includegraphics[width=.9\columnwidth]{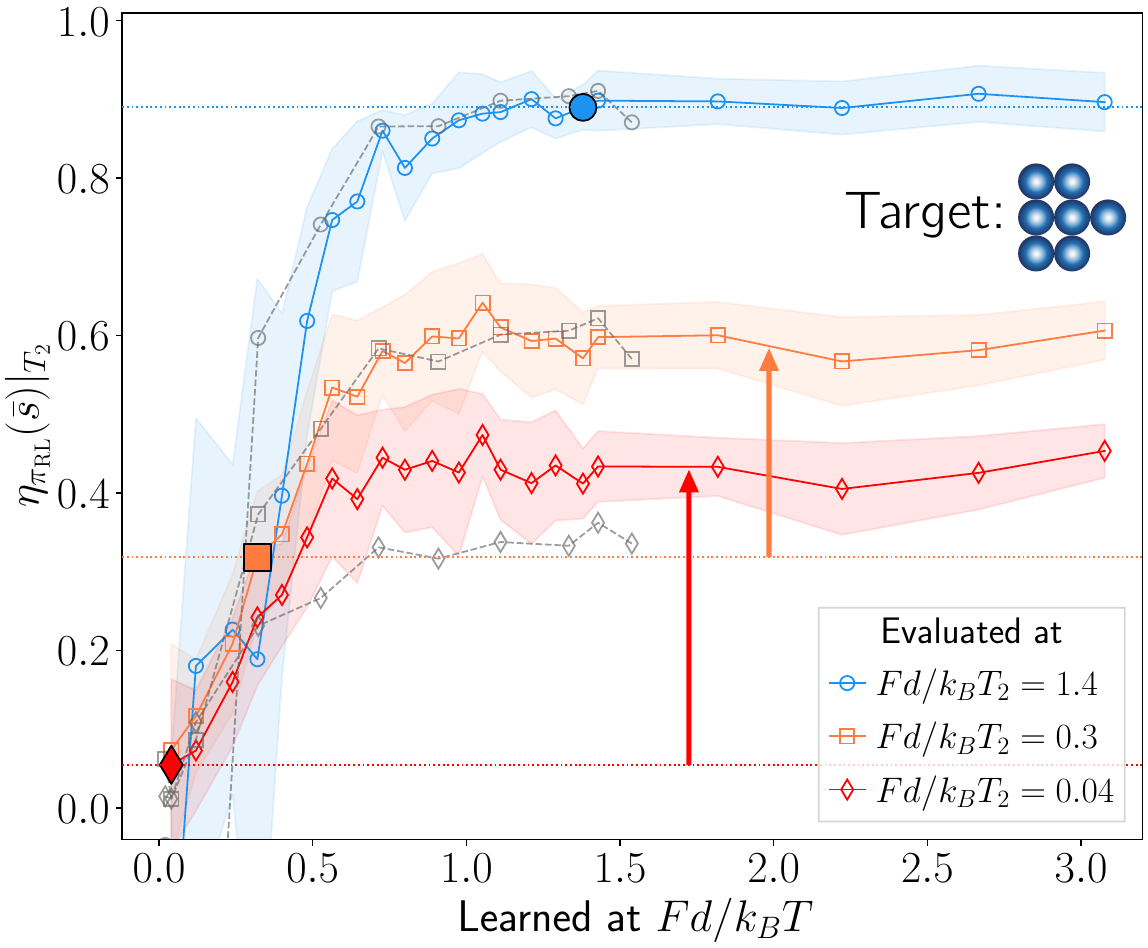}\hfill
    \caption{Transfer learning for a 7-particle target. For each curve, the value of $Fd/J$ is fixed : $Fd/J=0.4$ (continuous lines) and $Fd/J=0.2$ (dashed grey lines). The efficiency $\eta_{\pi_\text{RL}}(\bar{s})|_{T_2}$ is evaluated at a fixed temperature $T_2$ with policies $\pi_{RL}$ learned at various temperatures $T$ with the same value of $Fd/J$. For each point, we perform 10 independent learning runs. The average efficiency $\eta_{\pi_\text{RL}}(\bar{s})|_{T_2}$ as a function of $Fd/k_BT$ is shown with empty symbols. The width of the shaded area represents is the standard deviation. The efficiency increases and reaches a plateau  for low learning temperatures, roughly corresponding to $Fd/k_BT>1$. 
    The larger full symbols correspond to the average efficiency when learning and evaluation are done at the same temperature $T=T_2$. The arrows represent the gain of efficiency when learning at low temperatures $Fd/k_BT>1$. 
    }
    \label{fig:3}
\end{figure}

Remarkably, one can circumvent the drop of efficiency
at high temperature by learning at a lower temperature.
Indeed, policies learned at $k_BT\ll Fd$ can exhibit efficiencies up to $0.4$ 
when evaluated at a higher temperature $T_2$ such that $k_BT_2\gg Fd$,
while the efficiency would be vanishingly small if the policy was learned at $T_2$. 
In \cref{fig:3},
the efficiency of a policy $\pi_\text{RL}$ learned at a temperature $T$ and
evaluated at a temperature $T_2$ for different forces $F$ is seen
to depend only weakly on $J/k_BT$.
For $Fd/k_BT>1$, the efficiency saturates to a maximum value.
Note that learning at $Fd/k_BT>1$ means that the efficiency of learning is
good, i.e., that the learned policy is close to the optimal policy at the learning
temperature.
Such a transferability of the learned policy therefore suggests that there
is some persistence in the optimal policy when varying temperature.

We propose that this transferability is caused by the
tendency of the optimal policy close to the target to exhibit both a larger policy robustness 
and a larger relevance for the efficiency.
The larger relevance is due to the fact that one necessarily has
to go through these states close to the target to reach the target.
The larger robustness is caused by the fact that
the optimal choice of the force tends to be always directed towards the target
in states that are close to the target, with a small sensitivity to 
the values of the transition rates.
In { SM Sec. V}, we provide some numerical results
following hints from an expansion at small $Fd/k_BT$
that support these assumptions. The optimal policies at different temperatures 
are similar in states that are most relevant for the efficiency,
and these states are on average closer to the target (in terms of 
distance measured by the ring index~\cite{Baronchelli2006}).
Since the optimal actions in states that are close to the target are similar 
at high and low temperatures,
one can learn them at low temperatures where RL performs better,
and transfer them to higher temperatures.

%%%%%%%%%%%%%%%%%%%%%%%%%%%%%%%%%%%%%%%%%%%%%%%%%%%%%%%%%%%%%%
\paragraph{Discussion.}
%%%%%%%%%%%%%%%%%%%%%%%%%%%%%%%%%%%%%%%%%%%%%%%%%%%%%%%%%%%%%%

Our main result is that, due to thermal noise, Reinforcement Learning becomes inefficient
at high temperatures, small length-scales and small forces.
Hence, when downscaling devices toward the nano-scale, learning
becomes difficult or impossible. Our simulations also
show that, since the states that are
close to the target exhibit both a higher robustness and a larger relevance,
transfer learning is possible from conditions where learning can be achieved, such as lower temperatures.
This result is surprising since different regimes of fluctuation dynamics
are expected as the temperature is varied~\cite{Khare1995,PierreLouis2001,Giesen1995}.

The feasibility of some experimental RL control of colloidal clusters 
was demonstrated in Ref.~\cite{Zhang2020}, focusing on
relaxation of large clusters (of a few hundreds of particles)
towards a circular shape. Despite major differences
with our work---we consider arbitrary target shapes 
and few-particle clusters---we speculate that 
better learning at low temperatures and 
transferability of low-temperature policies to
high temperatures could be observed in this type of experimental system.

Furthermore, navigation policies for directed active colloids
were experimentally found to be transferable from one temperature to another in Ref.~\cite{MuinosLandin2021}.
We propose that this is caused by the persistence of the
optimal policy in high-sensitivity states. 
However, observations become more difficult at high temperatures due to faster particle motion.
Disentangling the observation-induced
failure of the learning process and the intrinsic difficulty
of learning at high temperatures is an open fundamental 
challenge that would require further theoretical developments.

Moreover, when noise is not too strong and when 
the systems or the learning algorithms are complex enough,
noise is not always detrimental and can help exploration
or can regularize the learning process~\cite{Igl2019,Arjovsky2017,Barfuss2023}.
However, our results, which have been checked here for MCL and QL,
only rely on the use of an $\varepsilon$-greedy policy based on the $q$ function in the limit of strong thermal noise.
We therefore expect them to be robust and relevant 
for a wider range of 
RL algorithms based on the $q$ function, including
advanced RL algorithms
that are coupled to artificial neural networks,
such as Deep Q-Learning~\cite{VanHasselt2016,Mnih2013}.
Hence, our study on elementary RL methods
provides directions for better performances 
and transfer learning strategies~\cite{Taylor2009,Zhuang2020} for advanced RL algorithms 
applied to nano-scale systems. 

\begin{acknowledgments}
The authors wish to thank Younes Benamara and Laetitia Matignon for useful discussions and David Rodney for comments on the manuscript.
\end{acknowledgments}

\bibliography{refs}

%apsrev4-2.bst 2019-01-14 (MD) hand-edited version of apsrev4-1.bst
%Control: key (0)
%Control: author (72) initials jnrlst
%Control: editor formatted (1) identically to author
%Control: production of article title (-1) disabled
%Control: page (0) single
%Control: year (1) truncated
%Control: production of eprint (0) enabled
\begin{thebibliography}{60}%
\makeatletter
\providecommand \@ifxundefined [1]{%
 \@ifx{#1\undefined}
}%
\providecommand \@ifnum [1]{%
 \ifnum #1\expandafter \@firstoftwo
 \else \expandafter \@secondoftwo
 \fi
}%
\providecommand \@ifx [1]{%
 \ifx #1\expandafter \@firstoftwo
 \else \expandafter \@secondoftwo
 \fi
}%
\providecommand \natexlab [1]{#1}%
\providecommand \enquote  [1]{``#1''}%
\providecommand \bibnamefont  [1]{#1}%
\providecommand \bibfnamefont [1]{#1}%
\providecommand \citenamefont [1]{#1}%
\providecommand \href@noop [0]{\@secondoftwo}%
\providecommand \href [0]{\begingroup \@sanitize@url \@href}%
\providecommand \@href[1]{\@@startlink{#1}\@@href}%
\providecommand \@@href[1]{\endgroup#1\@@endlink}%
\providecommand \@sanitize@url [0]{\catcode `\\12\catcode `\$12\catcode
  `\&12\catcode `\#12\catcode `\^12\catcode `\_12\catcode `\%12\relax}%
\providecommand \@@startlink[1]{}%
\providecommand \@@endlink[0]{}%
\providecommand \url  [0]{\begingroup\@sanitize@url \@url }%
\providecommand \@url [1]{\endgroup\@href {#1}{\urlprefix }}%
\providecommand \urlprefix  [0]{URL }%
\providecommand \Eprint [0]{\href }%
\providecommand \doibase [0]{https://doi.org/}%
\providecommand \selectlanguage [0]{\@gobble}%
\providecommand \bibinfo  [0]{\@secondoftwo}%
\providecommand \bibfield  [0]{\@secondoftwo}%
\providecommand \translation [1]{[#1]}%
\providecommand \BibitemOpen [0]{}%
\providecommand \bibitemStop [0]{}%
\providecommand \bibitemNoStop [0]{.\EOS\space}%
\providecommand \EOS [0]{\spacefactor3000\relax}%
\providecommand \BibitemShut  [1]{\csname bibitem#1\endcsname}%
\let\auto@bib@innerbib\@empty
%</preamble>
\bibitem [{\citenamefont {Sutton}\ and\ \citenamefont
  {Barto}(2018)}]{Sutton1998}%
  \BibitemOpen
  \bibfield  {author} {\bibinfo {author} {\bibfnamefont {R.~S.}\ \bibnamefont
  {Sutton}}\ and\ \bibinfo {author} {\bibfnamefont {A.~G.}\ \bibnamefont
  {Barto}},\ }\href {http://incompleteideas.net/book/the-book.html} {\emph
  {\bibinfo {title} {Reinforcement Learning: An Introduction}}},\ \bibinfo
  {edition} {2nd}\ ed.\ (\bibinfo  {publisher} {The MIT Press},\ \bibinfo
  {year} {2018})\BibitemShut {NoStop}%
\bibitem [{\citenamefont {Mnih}\ \emph {et~al.}(2013)\citenamefont {Mnih},
  \citenamefont {Kavukcuoglu}, \citenamefont {Silver}, \citenamefont {Graves},
  \citenamefont {Antonoglou}, \citenamefont {Wierstra},\ and\ \citenamefont
  {Riedmiller}}]{Mnih2013}%
  \BibitemOpen
  \bibfield  {author} {\bibinfo {author} {\bibfnamefont {V.}~\bibnamefont
  {Mnih}}, \bibinfo {author} {\bibfnamefont {K.}~\bibnamefont {Kavukcuoglu}},
  \bibinfo {author} {\bibfnamefont {D.}~\bibnamefont {Silver}}, \bibinfo
  {author} {\bibfnamefont {A.}~\bibnamefont {Graves}}, \bibinfo {author}
  {\bibfnamefont {I.}~\bibnamefont {Antonoglou}}, \bibinfo {author}
  {\bibfnamefont {D.}~\bibnamefont {Wierstra}},\ and\ \bibinfo {author}
  {\bibfnamefont {M.}~\bibnamefont {Riedmiller}},\ }\href@noop {} {\bibinfo
  {title} {Playing atari with deep reinforcement learning}} (\bibinfo {year}
  {2013}),\ \Eprint {https://arxiv.org/abs/1312.5602} {arXiv:1312.5602 [cs.LG]}
  \BibitemShut {NoStop}%
\bibitem [{\citenamefont {Kober}\ \emph {et~al.}(2013)\citenamefont {Kober},
  \citenamefont {Bagnell},\ and\ \citenamefont {Peters}}]{Kober2013}%
  \BibitemOpen
  \bibfield  {author} {\bibinfo {author} {\bibfnamefont {J.}~\bibnamefont
  {Kober}}, \bibinfo {author} {\bibfnamefont {J.~A.}\ \bibnamefont {Bagnell}},\
  and\ \bibinfo {author} {\bibfnamefont {J.}~\bibnamefont {Peters}},\ }\href
  {https://doi.org/10.1177/0278364913495721} {\bibfield  {journal} {\bibinfo
  {journal} {The International Journal of Robotics Research}\ }\textbf
  {\bibinfo {volume} {32}},\ \bibinfo {pages} {1238} (\bibinfo {year}
  {2013})},\ \Eprint
  {https://arxiv.org/abs/https://doi.org/10.1177/0278364913495721}
  {https://doi.org/10.1177/0278364913495721} \BibitemShut {NoStop}%
\bibitem [{\citenamefont {Zhao}\ \emph {et~al.}(2020)\citenamefont {Zhao},
  \citenamefont {Queralta},\ and\ \citenamefont {Westerlund}}]{Zhao2020}%
  \BibitemOpen
  \bibfield  {author} {\bibinfo {author} {\bibfnamefont {W.}~\bibnamefont
  {Zhao}}, \bibinfo {author} {\bibfnamefont {J.~P.}\ \bibnamefont {Queralta}},\
  and\ \bibinfo {author} {\bibfnamefont {T.}~\bibnamefont {Westerlund}},\ }in\
  \href {https://doi.org/10.1109/SSCI47803.2020.9308468} {\emph {\bibinfo
  {booktitle} {2020 IEEE Symposium Series on Computational Intelligence
  (SSCI)}}}\ (\bibinfo {year} {2020})\ pp.\ \bibinfo {pages}
  {737--744}\BibitemShut {NoStop}%
\bibitem [{\citenamefont {Garnier}\ \emph {et~al.}(2021)\citenamefont
  {Garnier}, \citenamefont {Viquerat}, \citenamefont {Rabault}, \citenamefont
  {Larcher}, \citenamefont {Kuhnle},\ and\ \citenamefont
  {Hachem}}]{Garnier2021}%
  \BibitemOpen
  \bibfield  {author} {\bibinfo {author} {\bibfnamefont {P.}~\bibnamefont
  {Garnier}}, \bibinfo {author} {\bibfnamefont {J.}~\bibnamefont {Viquerat}},
  \bibinfo {author} {\bibfnamefont {J.}~\bibnamefont {Rabault}}, \bibinfo
  {author} {\bibfnamefont {A.}~\bibnamefont {Larcher}}, \bibinfo {author}
  {\bibfnamefont {A.}~\bibnamefont {Kuhnle}},\ and\ \bibinfo {author}
  {\bibfnamefont {E.}~\bibnamefont {Hachem}},\ }\href
  {https://doi.org/https://doi.org/10.1016/j.compfluid.2021.104973} {\bibfield
  {journal} {\bibinfo  {journal} {Computers \& Fluids}\ }\textbf {\bibinfo
  {volume} {225}},\ \bibinfo {pages} {104973} (\bibinfo {year}
  {2021})}\BibitemShut {NoStop}%
\bibitem [{\citenamefont {Beintema}\ \emph {et~al.}(2020)\citenamefont
  {Beintema}, \citenamefont {Corbetta}, \citenamefont {Biferale},\ and\
  \citenamefont {Toschi}}]{Beintema2020}%
  \BibitemOpen
  \bibfield  {author} {\bibinfo {author} {\bibfnamefont {G.}~\bibnamefont
  {Beintema}}, \bibinfo {author} {\bibfnamefont {A.}~\bibnamefont {Corbetta}},
  \bibinfo {author} {\bibfnamefont {L.}~\bibnamefont {Biferale}},\ and\
  \bibinfo {author} {\bibfnamefont {F.}~\bibnamefont {Toschi}},\ }\href@noop {}
  {\bibfield  {journal} {\bibinfo  {journal} {Journal of Turbulence}\ }\textbf
  {\bibinfo {volume} {21}},\ \bibinfo {pages} {585} (\bibinfo {year}
  {2020})}\BibitemShut {NoStop}%
\bibitem [{\citenamefont {Singh}\ \emph {et~al.}(2023)\citenamefont {Singh},
  \citenamefont {van Breugel}, \citenamefont {Rao},\ and\ \citenamefont
  {Brunton}}]{Singh2023}%
  \BibitemOpen
  \bibfield  {author} {\bibinfo {author} {\bibfnamefont {S.~H.}\ \bibnamefont
  {Singh}}, \bibinfo {author} {\bibfnamefont {F.}~\bibnamefont {van Breugel}},
  \bibinfo {author} {\bibfnamefont {R.~P.}\ \bibnamefont {Rao}},\ and\ \bibinfo
  {author} {\bibfnamefont {B.~W.}\ \bibnamefont {Brunton}},\ }\href@noop {}
  {\bibfield  {journal} {\bibinfo  {journal} {Nature Machine Intelligence}\
  }\textbf {\bibinfo {volume} {5}},\ \bibinfo {pages} {58} (\bibinfo {year}
  {2023})}\BibitemShut {NoStop}%
\bibitem [{\citenamefont {Nasiri}\ and\ \citenamefont
  {Liebchen}(2022)}]{Nasiri2022}%
  \BibitemOpen
  \bibfield  {author} {\bibinfo {author} {\bibfnamefont {M.}~\bibnamefont
  {Nasiri}}\ and\ \bibinfo {author} {\bibfnamefont {B.}~\bibnamefont
  {Liebchen}},\ }\href@noop {} {\bibfield  {journal} {\bibinfo  {journal} {New
  Journal of Physics}\ }\textbf {\bibinfo {volume} {24}},\ \bibinfo {pages}
  {073042} (\bibinfo {year} {2022})}\BibitemShut {NoStop}%
\bibitem [{\citenamefont {Cichos}\ \emph {et~al.}(2020)\citenamefont {Cichos},
  \citenamefont {Gustavsson}, \citenamefont {Mehlig},\ and\ \citenamefont
  {Volpe}}]{Cichos2020}%
  \BibitemOpen
  \bibfield  {author} {\bibinfo {author} {\bibfnamefont {F.}~\bibnamefont
  {Cichos}}, \bibinfo {author} {\bibfnamefont {K.}~\bibnamefont {Gustavsson}},
  \bibinfo {author} {\bibfnamefont {B.}~\bibnamefont {Mehlig}},\ and\ \bibinfo
  {author} {\bibfnamefont {G.}~\bibnamefont {Volpe}},\ }\href
  {https://doi.org/10.1038/s42256-020-0146-9} {\bibfield  {journal} {\bibinfo
  {journal} {Nature Machine Intelligence}\ }\textbf {\bibinfo {volume} {2}},\
  \bibinfo {pages} {94} (\bibinfo {year} {2020})}\BibitemShut {NoStop}%
\bibitem [{\citenamefont {Yang}\ and\ \citenamefont {Bevan}(2020)}]{Yang2020}%
  \BibitemOpen
  \bibfield  {author} {\bibinfo {author} {\bibfnamefont {Y.}~\bibnamefont
  {Yang}}\ and\ \bibinfo {author} {\bibfnamefont {M.~A.}\ \bibnamefont
  {Bevan}},\ }\href {https://doi.org/10.1126/sciadv.aay7679} {\bibfield
  {journal} {\bibinfo  {journal} {Science Advances}\ }\textbf {\bibinfo
  {volume} {6}},\  (\bibinfo {year} {2020})}\BibitemShut {NoStop}%
\bibitem [{\citenamefont {Novati}\ \emph {et~al.}(2019)\citenamefont {Novati},
  \citenamefont {Mahadevan},\ and\ \citenamefont {Koumoutsakos}}]{Novati2019}%
  \BibitemOpen
  \bibfield  {author} {\bibinfo {author} {\bibfnamefont {G.}~\bibnamefont
  {Novati}}, \bibinfo {author} {\bibfnamefont {L.}~\bibnamefont {Mahadevan}},\
  and\ \bibinfo {author} {\bibfnamefont {P.}~\bibnamefont {Koumoutsakos}},\
  }\href@noop {} {\bibfield  {journal} {\bibinfo  {journal} {Physical Review
  Fluids}\ }\textbf {\bibinfo {volume} {4}},\ \bibinfo {pages} {093902}
  (\bibinfo {year} {2019})}\BibitemShut {NoStop}%
\bibitem [{\citenamefont {Schneider}\ and\ \citenamefont
  {Stark}(2019)}]{Schneider2019}%
  \BibitemOpen
  \bibfield  {author} {\bibinfo {author} {\bibfnamefont {E.}~\bibnamefont
  {Schneider}}\ and\ \bibinfo {author} {\bibfnamefont {H.}~\bibnamefont
  {Stark}},\ }\href@noop {} {\bibfield  {journal} {\bibinfo  {journal}
  {Europhysics Letters}\ }\textbf {\bibinfo {volume} {127}},\ \bibinfo {pages}
  {64003} (\bibinfo {year} {2019})}\BibitemShut {NoStop}%
\bibitem [{\citenamefont {Reddy}\ \emph {et~al.}(2018)\citenamefont {Reddy},
  \citenamefont {Wong-Ng}, \citenamefont {Celani}, \citenamefont {Sejnowski},\
  and\ \citenamefont {Vergassola}}]{Reddy2018}%
  \BibitemOpen
  \bibfield  {author} {\bibinfo {author} {\bibfnamefont {G.}~\bibnamefont
  {Reddy}}, \bibinfo {author} {\bibfnamefont {J.}~\bibnamefont {Wong-Ng}},
  \bibinfo {author} {\bibfnamefont {A.}~\bibnamefont {Celani}}, \bibinfo
  {author} {\bibfnamefont {T.~J.}\ \bibnamefont {Sejnowski}},\ and\ \bibinfo
  {author} {\bibfnamefont {M.}~\bibnamefont {Vergassola}},\ }\href@noop {}
  {\bibfield  {journal} {\bibinfo  {journal} {Nature}\ }\textbf {\bibinfo
  {volume} {562}},\ \bibinfo {pages} {236} (\bibinfo {year}
  {2018})}\BibitemShut {NoStop}%
\bibitem [{\citenamefont {Reddy}\ \emph {et~al.}(2016)\citenamefont {Reddy},
  \citenamefont {Celani}, \citenamefont {Sejnowski},\ and\ \citenamefont
  {Vergassola}}]{Reddy2016}%
  \BibitemOpen
  \bibfield  {author} {\bibinfo {author} {\bibfnamefont {G.}~\bibnamefont
  {Reddy}}, \bibinfo {author} {\bibfnamefont {A.}~\bibnamefont {Celani}},
  \bibinfo {author} {\bibfnamefont {T.~J.}\ \bibnamefont {Sejnowski}},\ and\
  \bibinfo {author} {\bibfnamefont {M.}~\bibnamefont {Vergassola}},\ }\href
  {https://doi.org/10.1073/pnas.1606075113} {\bibfield  {journal} {\bibinfo
  {journal} {Proceedings of the National Academy of Sciences}\ }\textbf
  {\bibinfo {volume} {113}},\  (\bibinfo {year} {2016})}\BibitemShut {NoStop}%
\bibitem [{\citenamefont {Colabrese}\ \emph {et~al.}(2017)\citenamefont
  {Colabrese}, \citenamefont {Gustavsson}, \citenamefont {Celani},\ and\
  \citenamefont {Biferale}}]{Colabrese2017}%
  \BibitemOpen
  \bibfield  {author} {\bibinfo {author} {\bibfnamefont {S.}~\bibnamefont
  {Colabrese}}, \bibinfo {author} {\bibfnamefont {K.}~\bibnamefont
  {Gustavsson}}, \bibinfo {author} {\bibfnamefont {A.}~\bibnamefont {Celani}},\
  and\ \bibinfo {author} {\bibfnamefont {L.}~\bibnamefont {Biferale}},\ }\href
  {https://doi.org/10.1103/PhysRevLett.118.158004} {\bibfield  {journal}
  {\bibinfo  {journal} {Phys. Rev. Lett.}\ }\textbf {\bibinfo {volume} {118}},\
  \bibinfo {pages} {158004} (\bibinfo {year} {2017})}\BibitemShut {NoStop}%
\bibitem [{\citenamefont {Durve}\ \emph {et~al.}(2020)\citenamefont {Durve},
  \citenamefont {Peruani},\ and\ \citenamefont {Celani}}]{Durve2020}%
  \BibitemOpen
  \bibfield  {author} {\bibinfo {author} {\bibfnamefont {M.}~\bibnamefont
  {Durve}}, \bibinfo {author} {\bibfnamefont {F.}~\bibnamefont {Peruani}},\
  and\ \bibinfo {author} {\bibfnamefont {A.}~\bibnamefont {Celani}},\ }\href
  {https://doi.org/10.1103/PhysRevE.102.012601} {\bibfield  {journal} {\bibinfo
   {journal} {Phys. Rev. E}\ }\textbf {\bibinfo {volume} {102}},\ \bibinfo
  {pages} {012601} (\bibinfo {year} {2020})}\BibitemShut {NoStop}%
\bibitem [{\citenamefont {Mui{\~{n}}os-Landin}\ \emph
  {et~al.}(2021)\citenamefont {Mui{\~{n}}os-Landin}, \citenamefont {Fischer},
  \citenamefont {Holubec},\ and\ \citenamefont {Cichos}}]{MuinosLandin2021}%
  \BibitemOpen
  \bibfield  {author} {\bibinfo {author} {\bibfnamefont {S.}~\bibnamefont
  {Mui{\~{n}}os-Landin}}, \bibinfo {author} {\bibfnamefont {A.}~\bibnamefont
  {Fischer}}, \bibinfo {author} {\bibfnamefont {V.}~\bibnamefont {Holubec}},\
  and\ \bibinfo {author} {\bibfnamefont {F.}~\bibnamefont {Cichos}},\ }\href
  {https://doi.org/10.1126/scirobotics.abd9285} {\bibfield  {journal} {\bibinfo
   {journal} {Science Robotics}\ }\textbf {\bibinfo {volume} {6}},\  (\bibinfo
  {year} {2021})}\BibitemShut {NoStop}%
\bibitem [{\citenamefont {Zhang}\ \emph {et~al.}(2020)\citenamefont {Zhang},
  \citenamefont {Yang}, \citenamefont {Zhang},\ and\ \citenamefont
  {Bevan}}]{Zhang2020}%
  \BibitemOpen
  \bibfield  {author} {\bibinfo {author} {\bibfnamefont {J.}~\bibnamefont
  {Zhang}}, \bibinfo {author} {\bibfnamefont {J.}~\bibnamefont {Yang}},
  \bibinfo {author} {\bibfnamefont {Y.}~\bibnamefont {Zhang}},\ and\ \bibinfo
  {author} {\bibfnamefont {M.~A.}\ \bibnamefont {Bevan}},\ }\href
  {https://doi.org/10.1126/sciadv.abd6716} {\bibfield  {journal} {\bibinfo
  {journal} {Science Advances}\ }\textbf {\bibinfo {volume} {6}},\  (\bibinfo
  {year} {2020})}\BibitemShut {NoStop}%
\bibitem [{\citenamefont {Fletcher}\ \emph {et~al.}(2005)\citenamefont
  {Fletcher}, \citenamefont {Dumur}, \citenamefont {Pollard},\ and\
  \citenamefont {Feringa}}]{Fletcher2005}%
  \BibitemOpen
  \bibfield  {author} {\bibinfo {author} {\bibfnamefont {S.~P.}\ \bibnamefont
  {Fletcher}}, \bibinfo {author} {\bibfnamefont {F.}~\bibnamefont {Dumur}},
  \bibinfo {author} {\bibfnamefont {M.~M.}\ \bibnamefont {Pollard}},\ and\
  \bibinfo {author} {\bibfnamefont {B.~L.}\ \bibnamefont {Feringa}},\
  }\href@noop {} {\bibfield  {journal} {\bibinfo  {journal} {Science}\ }\textbf
  {\bibinfo {volume} {310}},\ \bibinfo {pages} {80} (\bibinfo {year}
  {2005})}\BibitemShut {NoStop}%
\bibitem [{\citenamefont {Mallouk}\ and\ \citenamefont
  {Sen}(2009)}]{Mallouk2009}%
  \BibitemOpen
  \bibfield  {author} {\bibinfo {author} {\bibfnamefont {T.~E.}\ \bibnamefont
  {Mallouk}}\ and\ \bibinfo {author} {\bibfnamefont {A.}~\bibnamefont {Sen}},\
  }\href@noop {} {\bibfield  {journal} {\bibinfo  {journal} {Scientific
  American}\ }\textbf {\bibinfo {volume} {300}},\ \bibinfo {pages} {72}
  (\bibinfo {year} {2009})}\BibitemShut {NoStop}%
\bibitem [{\citenamefont {Shao}\ and\ \citenamefont
  {K{\"a}ll}(2018)}]{Shao2018}%
  \BibitemOpen
  \bibfield  {author} {\bibinfo {author} {\bibfnamefont {L.}~\bibnamefont
  {Shao}}\ and\ \bibinfo {author} {\bibfnamefont {M.}~\bibnamefont
  {K{\"a}ll}},\ }\href@noop {} {\bibfield  {journal} {\bibinfo  {journal}
  {Advanced Functional Materials}\ }\textbf {\bibinfo {volume} {28}},\ \bibinfo
  {pages} {1706272} (\bibinfo {year} {2018})}\BibitemShut {NoStop}%
\bibitem [{\citenamefont {Robertson}\ \emph {et~al.}(2018)\citenamefont
  {Robertson}, \citenamefont {Huang}, \citenamefont {Chen},\ and\ \citenamefont
  {Kapral}}]{Robertson2018}%
  \BibitemOpen
  \bibfield  {author} {\bibinfo {author} {\bibfnamefont {B.}~\bibnamefont
  {Robertson}}, \bibinfo {author} {\bibfnamefont {M.-J.}\ \bibnamefont
  {Huang}}, \bibinfo {author} {\bibfnamefont {J.-X.}\ \bibnamefont {Chen}},\
  and\ \bibinfo {author} {\bibfnamefont {R.}~\bibnamefont {Kapral}},\
  }\href@noop {} {\bibfield  {journal} {\bibinfo  {journal} {Accounts of
  Chemical Research}\ }\textbf {\bibinfo {volume} {51}},\ \bibinfo {pages}
  {2355} (\bibinfo {year} {2018})}\BibitemShut {NoStop}%
\bibitem [{\citenamefont {Zhang}\ \emph {et~al.}(2022)\citenamefont {Zhang},
  \citenamefont {Li}, \citenamefont {Duan}, \citenamefont {Abbas},
  \citenamefont {Mundaca-Uribe}, \citenamefont {Yin}, \citenamefont {Luan},
  \citenamefont {Gao}, \citenamefont {Fang}, \citenamefont {Zhang},\ and\
  \citenamefont {Wang}}]{Zhang2022}%
  \BibitemOpen
  \bibfield  {author} {\bibinfo {author} {\bibfnamefont {F.}~\bibnamefont
  {Zhang}}, \bibinfo {author} {\bibfnamefont {Z.}~\bibnamefont {Li}}, \bibinfo
  {author} {\bibfnamefont {Y.}~\bibnamefont {Duan}}, \bibinfo {author}
  {\bibfnamefont {A.}~\bibnamefont {Abbas}}, \bibinfo {author} {\bibfnamefont
  {R.}~\bibnamefont {Mundaca-Uribe}}, \bibinfo {author} {\bibfnamefont
  {L.}~\bibnamefont {Yin}}, \bibinfo {author} {\bibfnamefont {H.}~\bibnamefont
  {Luan}}, \bibinfo {author} {\bibfnamefont {W.}~\bibnamefont {Gao}}, \bibinfo
  {author} {\bibfnamefont {R.~H.}\ \bibnamefont {Fang}}, \bibinfo {author}
  {\bibfnamefont {L.}~\bibnamefont {Zhang}},\ and\ \bibinfo {author}
  {\bibfnamefont {J.}~\bibnamefont {Wang}},\ }\href
  {https://doi.org/10.1126/scirobotics.abo4160} {\bibfield  {journal} {\bibinfo
   {journal} {Science Robotics}\ }\textbf {\bibinfo {volume} {7}},\  (\bibinfo
  {year} {2022})}\BibitemShut {NoStop}%
\bibitem [{\citenamefont {Li}\ \emph {et~al.}(2016)\citenamefont {Li},
  \citenamefont {Rozen},\ and\ \citenamefont {Wang}}]{Li2016}%
  \BibitemOpen
  \bibfield  {author} {\bibinfo {author} {\bibfnamefont {J.}~\bibnamefont
  {Li}}, \bibinfo {author} {\bibfnamefont {I.}~\bibnamefont {Rozen}},\ and\
  \bibinfo {author} {\bibfnamefont {J.}~\bibnamefont {Wang}},\ }\href@noop {}
  {\bibfield  {journal} {\bibinfo  {journal} {ACS nano}\ }\textbf {\bibinfo
  {volume} {10}},\ \bibinfo {pages} {5619} (\bibinfo {year}
  {2016})}\BibitemShut {NoStop}%
\bibitem [{\citenamefont {Li}\ \emph {et~al.}(2018)\citenamefont {Li},
  \citenamefont {Jiang}, \citenamefont {Liu}, \citenamefont {Zhang},
  \citenamefont {Tian}, \citenamefont {Song}, \citenamefont {Wang},
  \citenamefont {Zou}, \citenamefont {Anderson}, \citenamefont {Han} \emph
  {et~al.}}]{Li2018}%
  \BibitemOpen
  \bibfield  {author} {\bibinfo {author} {\bibfnamefont {S.}~\bibnamefont
  {Li}}, \bibinfo {author} {\bibfnamefont {Q.}~\bibnamefont {Jiang}}, \bibinfo
  {author} {\bibfnamefont {S.}~\bibnamefont {Liu}}, \bibinfo {author}
  {\bibfnamefont {Y.}~\bibnamefont {Zhang}}, \bibinfo {author} {\bibfnamefont
  {Y.}~\bibnamefont {Tian}}, \bibinfo {author} {\bibfnamefont {C.}~\bibnamefont
  {Song}}, \bibinfo {author} {\bibfnamefont {J.}~\bibnamefont {Wang}}, \bibinfo
  {author} {\bibfnamefont {Y.}~\bibnamefont {Zou}}, \bibinfo {author}
  {\bibfnamefont {G.~J.}\ \bibnamefont {Anderson}}, \bibinfo {author}
  {\bibfnamefont {J.-Y.}\ \bibnamefont {Han}}, \emph {et~al.},\ }\href@noop {}
  {\bibfield  {journal} {\bibinfo  {journal} {Nature biotechnology}\ }\textbf
  {\bibinfo {volume} {36}},\ \bibinfo {pages} {258} (\bibinfo {year}
  {2018})}\BibitemShut {NoStop}%
\bibitem [{\citenamefont {Zarei}\ and\ \citenamefont
  {Zarei}(2018)}]{Zarei2018}%
  \BibitemOpen
  \bibfield  {author} {\bibinfo {author} {\bibfnamefont {M.}~\bibnamefont
  {Zarei}}\ and\ \bibinfo {author} {\bibfnamefont {M.}~\bibnamefont {Zarei}},\
  }\href@noop {} {\bibfield  {journal} {\bibinfo  {journal} {Small}\ }\textbf
  {\bibinfo {volume} {14}},\ \bibinfo {pages} {1800912} (\bibinfo {year}
  {2018})}\BibitemShut {NoStop}%
\bibitem [{\citenamefont {Nilim}\ and\ \citenamefont
  {El~Ghaoui}(2005)}]{Nilim2005}%
  \BibitemOpen
  \bibfield  {author} {\bibinfo {author} {\bibfnamefont {A.}~\bibnamefont
  {Nilim}}\ and\ \bibinfo {author} {\bibfnamefont {L.}~\bibnamefont
  {El~Ghaoui}},\ }\href@noop {} {\bibfield  {journal} {\bibinfo  {journal}
  {Operations Research}\ }\textbf {\bibinfo {volume} {53}},\ \bibinfo {pages}
  {780} (\bibinfo {year} {2005})}\BibitemShut {NoStop}%
\bibitem [{\citenamefont {Iyengar}(2005)}]{Iyengar2005}%
  \BibitemOpen
  \bibfield  {author} {\bibinfo {author} {\bibfnamefont {G.~N.}\ \bibnamefont
  {Iyengar}},\ }\href@noop {} {\bibfield  {journal} {\bibinfo  {journal}
  {Mathematics of Operations Research}\ }\textbf {\bibinfo {volume} {30}},\
  \bibinfo {pages} {257} (\bibinfo {year} {2005})}\BibitemShut {NoStop}%
\bibitem [{\citenamefont {Wang}\ \emph {et~al.}(2020)\citenamefont {Wang},
  \citenamefont {Liu},\ and\ \citenamefont {Li}}]{Wang2020}%
  \BibitemOpen
  \bibfield  {author} {\bibinfo {author} {\bibfnamefont {J.}~\bibnamefont
  {Wang}}, \bibinfo {author} {\bibfnamefont {Y.}~\bibnamefont {Liu}},\ and\
  \bibinfo {author} {\bibfnamefont {B.}~\bibnamefont {Li}},\ }\href
  {https://doi.org/10.1609/aaai.v34i04.6086} {\bibfield  {journal} {\bibinfo
  {journal} {Proceedings of the {AAAI} Conference on Artificial Intelligence}\
  }\textbf {\bibinfo {volume} {34}},\ \bibinfo {pages} {6202} (\bibinfo {year}
  {2020})}\BibitemShut {NoStop}%
\bibitem [{\citenamefont {Brunke}\ \emph {et~al.}(2022)\citenamefont {Brunke},
  \citenamefont {Greeff}, \citenamefont {Hall}, \citenamefont {Yuan},
  \citenamefont {Zhou}, \citenamefont {Panerati},\ and\ \citenamefont
  {Schoellig}}]{Brunke2022}%
  \BibitemOpen
  \bibfield  {author} {\bibinfo {author} {\bibfnamefont {L.}~\bibnamefont
  {Brunke}}, \bibinfo {author} {\bibfnamefont {M.}~\bibnamefont {Greeff}},
  \bibinfo {author} {\bibfnamefont {A.~W.}\ \bibnamefont {Hall}}, \bibinfo
  {author} {\bibfnamefont {Z.}~\bibnamefont {Yuan}}, \bibinfo {author}
  {\bibfnamefont {S.}~\bibnamefont {Zhou}}, \bibinfo {author} {\bibfnamefont
  {J.}~\bibnamefont {Panerati}},\ and\ \bibinfo {author} {\bibfnamefont
  {A.~P.}\ \bibnamefont {Schoellig}},\ }\href
  {https://doi.org/10.1146/annurev-control-042920-020211} {\bibfield  {journal}
  {\bibinfo  {journal} {Annual Review of Control, Robotics, and Autonomous
  Systems}\ }\textbf {\bibinfo {volume} {5}},\ \bibinfo {pages} {411} (\bibinfo
  {year} {2022})},\ \Eprint
  {https://arxiv.org/abs/https://doi.org/10.1146/annurev-control-042920-020211}
  {https://doi.org/10.1146/annurev-control-042920-020211} \BibitemShut
  {NoStop}%
\bibitem [{\citenamefont {Dridi}\ and\ \citenamefont
  {Lehmann}(2015)}]{Dridi2015}%
  \BibitemOpen
  \bibfield  {author} {\bibinfo {author} {\bibfnamefont {S.}~\bibnamefont
  {Dridi}}\ and\ \bibinfo {author} {\bibfnamefont {L.}~\bibnamefont
  {Lehmann}},\ }\href
  {https://doi.org/https://doi.org/10.1016/j.anbehav.2015.01.037} {\bibfield
  {journal} {\bibinfo  {journal} {Animal Behaviour}\ }\textbf {\bibinfo
  {volume} {104}},\ \bibinfo {pages} {87} (\bibinfo {year} {2015})}\BibitemShut
  {NoStop}%
\bibitem [{\citenamefont {Larchenko}\ \emph {et~al.}(2021)\citenamefont
  {Larchenko}, \citenamefont {Osinenko}, \citenamefont {Yaremenko},\ and\
  \citenamefont {Palyulin}}]{Larchenko2021}%
  \BibitemOpen
  \bibfield  {author} {\bibinfo {author} {\bibfnamefont {M.~A.}\ \bibnamefont
  {Larchenko}}, \bibinfo {author} {\bibfnamefont {P.}~\bibnamefont {Osinenko}},
  \bibinfo {author} {\bibfnamefont {G.}~\bibnamefont {Yaremenko}},\ and\
  \bibinfo {author} {\bibfnamefont {V.~V.}\ \bibnamefont {Palyulin}},\ }\href
  {https://doi.org/10.1109/ACCESS.2021.3129709} {\bibfield  {journal} {\bibinfo
   {journal} {IEEE Access}\ }\textbf {\bibinfo {volume} {9}},\ \bibinfo {pages}
  {159349} (\bibinfo {year} {2021})}\BibitemShut {NoStop}%
\bibitem [{\citenamefont {Boccardo}\ and\ \citenamefont
  {Pierre-Louis}(2022{\natexlab{a}})}]{Boccardo2022}%
  \BibitemOpen
  \bibfield  {author} {\bibinfo {author} {\bibfnamefont {F.}~\bibnamefont
  {Boccardo}}\ and\ \bibinfo {author} {\bibfnamefont {O.}~\bibnamefont
  {Pierre-Louis}},\ }\href {https://doi.org/10.1103/PhysRevLett.128.256102}
  {\bibfield  {journal} {\bibinfo  {journal} {Phys. Rev. Lett.}\ }\textbf
  {\bibinfo {volume} {128}},\ \bibinfo {pages} {256102} (\bibinfo {year}
  {2022}{\natexlab{a}})}\BibitemShut {NoStop}%
\bibitem [{\citenamefont {Kuhn}\ \emph {et~al.}(2005)\citenamefont {Kuhn},
  \citenamefont {Krug}, \citenamefont {Hausser},\ and\ \citenamefont
  {Voigt}}]{Kuhn2005}%
  \BibitemOpen
  \bibfield  {author} {\bibinfo {author} {\bibfnamefont {P.}~\bibnamefont
  {Kuhn}}, \bibinfo {author} {\bibfnamefont {J.}~\bibnamefont {Krug}}, \bibinfo
  {author} {\bibfnamefont {F.}~\bibnamefont {Hausser}},\ and\ \bibinfo {author}
  {\bibfnamefont {A.}~\bibnamefont {Voigt}},\ }\href
  {https://doi.org/10.1103/PhysRevLett.94.166105} {\bibfield  {journal}
  {\bibinfo  {journal} {Phys. Rev. Lett.}\ }\textbf {\bibinfo {volume} {94}},\
  \bibinfo {pages} {166105} (\bibinfo {year} {2005})}\BibitemShut {NoStop}%
\bibitem [{\citenamefont {Mahadevan}\ and\ \citenamefont
  {Bradley}(1999)}]{Mahadevan1999}%
  \BibitemOpen
  \bibfield  {author} {\bibinfo {author} {\bibfnamefont {M.}~\bibnamefont
  {Mahadevan}}\ and\ \bibinfo {author} {\bibfnamefont {R.~M.}\ \bibnamefont
  {Bradley}},\ }\href {https://doi.org/10.1103/PhysRevB.59.11037} {\bibfield
  {journal} {\bibinfo  {journal} {Phys. Rev. B}\ }\textbf {\bibinfo {volume}
  {59}},\ \bibinfo {pages} {11037} (\bibinfo {year} {1999})}\BibitemShut
  {NoStop}%
\bibitem [{\citenamefont {Pierre-Louis}\ and\ \citenamefont
  {Einstein}(2000)}]{PierreLouis2000}%
  \BibitemOpen
  \bibfield  {author} {\bibinfo {author} {\bibfnamefont {O.}~\bibnamefont
  {Pierre-Louis}}\ and\ \bibinfo {author} {\bibfnamefont {T.~L.}\ \bibnamefont
  {Einstein}},\ }\href {https://doi.org/10.1103/physrevb.62.13697} {\bibfield
  {journal} {\bibinfo  {journal} {Phys. Rev. B}\ }\textbf {\bibinfo {volume}
  {62}},\ \bibinfo {pages} {13697} (\bibinfo {year} {2000})}\BibitemShut
  {NoStop}%
\bibitem [{\citenamefont {Curiotto}\ \emph {et~al.}(2019)\citenamefont
  {Curiotto}, \citenamefont {Leroy}, \citenamefont {M\"{u}ller}, \citenamefont
  {Cheynis}, \citenamefont {Michailov}, \citenamefont {El-Barraj},\ and\
  \citenamefont {Ranguelov}}]{Curiotto2019}%
  \BibitemOpen
  \bibfield  {author} {\bibinfo {author} {\bibfnamefont {S.}~\bibnamefont
  {Curiotto}}, \bibinfo {author} {\bibfnamefont {F.}~\bibnamefont {Leroy}},
  \bibinfo {author} {\bibfnamefont {P.}~\bibnamefont {M\"{u}ller}}, \bibinfo
  {author} {\bibfnamefont {F.}~\bibnamefont {Cheynis}}, \bibinfo {author}
  {\bibfnamefont {M.}~\bibnamefont {Michailov}}, \bibinfo {author}
  {\bibfnamefont {A.}~\bibnamefont {El-Barraj}},\ and\ \bibinfo {author}
  {\bibfnamefont {B.}~\bibnamefont {Ranguelov}},\ }\href
  {https://doi.org/10.1016/j.jcrysgro.2019.05.016} {\bibfield  {journal}
  {\bibinfo  {journal} {J. Cryst. Growth}\ }\textbf {\bibinfo {volume} {520}},\
  \bibinfo {pages} {42} (\bibinfo {year} {2019})}\BibitemShut {NoStop}%
\bibitem [{\citenamefont {McCormack}\ \emph {et~al.}(2018)\citenamefont
  {McCormack}, \citenamefont {Han},\ and\ \citenamefont {Yan}}]{McCormack2018}%
  \BibitemOpen
  \bibfield  {author} {\bibinfo {author} {\bibfnamefont {P.}~\bibnamefont
  {McCormack}}, \bibinfo {author} {\bibfnamefont {F.}~\bibnamefont {Han}},\
  and\ \bibinfo {author} {\bibfnamefont {Z.}~\bibnamefont {Yan}},\ }\href
  {https://doi.org/10.1021/acs.jpclett.7b03188} {\bibfield  {journal} {\bibinfo
   {journal} {J. Phys. Chem. Lett.}\ }\textbf {\bibinfo {volume} {9}},\
  \bibinfo {pages} {545} (\bibinfo {year} {2018})}\BibitemShut {NoStop}%
\bibitem [{\citenamefont {Xue}\ \emph {et~al.}(2014)\citenamefont {Xue},
  \citenamefont {Beltran-Villegas}, \citenamefont {Tang}, \citenamefont
  {Bevan},\ and\ \citenamefont {Grover}}]{Xue2014}%
  \BibitemOpen
  \bibfield  {author} {\bibinfo {author} {\bibfnamefont {Y.}~\bibnamefont
  {Xue}}, \bibinfo {author} {\bibfnamefont {D.~J.}\ \bibnamefont
  {Beltran-Villegas}}, \bibinfo {author} {\bibfnamefont {X.}~\bibnamefont
  {Tang}}, \bibinfo {author} {\bibfnamefont {M.~A.}\ \bibnamefont {Bevan}},\
  and\ \bibinfo {author} {\bibfnamefont {M.~A.}\ \bibnamefont {Grover}},\
  }\href {https://doi.org/10.1109/tcst.2013.2296700} {\bibfield  {journal}
  {\bibinfo  {journal} {IEEE Trans. Control Syst. Technol.}\ }\textbf {\bibinfo
  {volume} {22}},\ \bibinfo {pages} {1956} (\bibinfo {year}
  {2014})}\BibitemShut {NoStop}%
\bibitem [{\citenamefont {Xue}\ \emph {et~al.}(2013)\citenamefont {Xue},
  \citenamefont {Beltran-Villegas}, \citenamefont {Bevan},\ and\ \citenamefont
  {Grover}}]{Xue2013}%
  \BibitemOpen
  \bibfield  {author} {\bibinfo {author} {\bibfnamefont {Y.}~\bibnamefont
  {Xue}}, \bibinfo {author} {\bibfnamefont {D.~J.}\ \bibnamefont
  {Beltran-Villegas}}, \bibinfo {author} {\bibfnamefont {M.~A.}\ \bibnamefont
  {Bevan}},\ and\ \bibinfo {author} {\bibfnamefont {M.~A.}\ \bibnamefont
  {Grover}},\ }in\ \href@noop {} {\emph {\bibinfo {booktitle} {2013 American
  Control Conference}}}\ (\bibinfo {organization} {IEEE},\ \bibinfo {year}
  {2013})\ pp.\ \bibinfo {pages} {3397--3402}\BibitemShut {NoStop}%
\bibitem [{\citenamefont {Banerjee}\ \emph {et~al.}(2009)\citenamefont
  {Banerjee}, \citenamefont {Pomerance}, \citenamefont {Losert},\ and\
  \citenamefont {Gupta}}]{Banerjee2009}%
  \BibitemOpen
  \bibfield  {author} {\bibinfo {author} {\bibfnamefont {A.~G.}\ \bibnamefont
  {Banerjee}}, \bibinfo {author} {\bibfnamefont {A.}~\bibnamefont {Pomerance}},
  \bibinfo {author} {\bibfnamefont {W.}~\bibnamefont {Losert}},\ and\ \bibinfo
  {author} {\bibfnamefont {S.~K.}\ \bibnamefont {Gupta}},\ }\href@noop {}
  {\bibfield  {journal} {\bibinfo  {journal} {IEEE Transactions on automation
  science and engineering}\ }\textbf {\bibinfo {volume} {7}},\ \bibinfo {pages}
  {218} (\bibinfo {year} {2009})}\BibitemShut {NoStop}%
\bibitem [{\citenamefont {Glasstone}\ \emph {et~al.}(1941)\citenamefont
  {Glasstone}, \citenamefont {Laidler},\ and\ \citenamefont
  {Eyring}}]{Glasstone1941}%
  \BibitemOpen
  \bibfield  {author} {\bibinfo {author} {\bibfnamefont {S.}~\bibnamefont
  {Glasstone}}, \bibinfo {author} {\bibfnamefont {K.~J.}\ \bibnamefont
  {Laidler}},\ and\ \bibinfo {author} {\bibfnamefont {H.}~\bibnamefont
  {Eyring}},\ }\href@noop {} {\emph {\bibinfo {title} {The Theory of Rate
  Processes: The Kinetics of Chemical Reactions, Viscosity, Diffusion and
  Electrochemical Phenomena}}},\ International chemical series\ (\bibinfo
  {publisher} {McGraw-Hill Book Company, Inc.},\ \bibinfo {year}
  {1941})\BibitemShut {NoStop}%
\bibitem [{\citenamefont {Liu}\ and\ \citenamefont {Weeks}(1998)}]{Liu1998}%
  \BibitemOpen
  \bibfield  {author} {\bibinfo {author} {\bibfnamefont {D.-J.}\ \bibnamefont
  {Liu}}\ and\ \bibinfo {author} {\bibfnamefont {J.~D.}\ \bibnamefont
  {Weeks}},\ }\href {https://doi.org/10.1103/PhysRevB.57.14891} {\bibfield
  {journal} {\bibinfo  {journal} {Phys. Rev. B}\ }\textbf {\bibinfo {volume}
  {57}},\ \bibinfo {pages} {14891} (\bibinfo {year} {1998})}\BibitemShut
  {NoStop}%
\bibitem [{\citenamefont {Giesen}(2001)}]{Giesen2001}%
  \BibitemOpen
  \bibfield  {author} {\bibinfo {author} {\bibfnamefont {M.}~\bibnamefont
  {Giesen}},\ }\href {https://doi.org/10.1016/s0079-6816(00)00021-6} {\bibfield
   {journal} {\bibinfo  {journal} {Prog. Surf. Sci.}\ }\textbf {\bibinfo
  {volume} {68}},\ \bibinfo {pages} {1} (\bibinfo {year} {2001})}\BibitemShut
  {NoStop}%
\bibitem [{\citenamefont {Tao}\ \emph {et~al.}(2010)\citenamefont {Tao},
  \citenamefont {Cullen},\ and\ \citenamefont {Williams}}]{Tao2010}%
  \BibitemOpen
  \bibfield  {author} {\bibinfo {author} {\bibfnamefont {C.}~\bibnamefont
  {Tao}}, \bibinfo {author} {\bibfnamefont {W.~G.}\ \bibnamefont {Cullen}},\
  and\ \bibinfo {author} {\bibfnamefont {E.~D.}\ \bibnamefont {Williams}},\
  }\href {https://doi.org/10.1126/science.1186648} {\bibfield  {journal}
  {\bibinfo  {journal} {Science}\ }\textbf {\bibinfo {volume} {328}},\ \bibinfo
  {pages} {736} (\bibinfo {year} {2010})}\BibitemShut {NoStop}%
\bibitem [{\citenamefont {Hubartt}\ and\ \citenamefont
  {Amar}(2015)}]{Hubartt2015}%
  \BibitemOpen
  \bibfield  {author} {\bibinfo {author} {\bibfnamefont {B.~C.}\ \bibnamefont
  {Hubartt}}\ and\ \bibinfo {author} {\bibfnamefont {J.~G.}\ \bibnamefont
  {Amar}},\ }\href@noop {} {\bibfield  {journal} {\bibinfo  {journal} {J. Chem.
  Phys.}\ }\textbf {\bibinfo {volume} {142}},\ \bibinfo {pages} {024709}
  (\bibinfo {year} {2015})}\BibitemShut {NoStop}%
\bibitem [{\citenamefont {Boccardo}\ \emph {et~al.}(2022)\citenamefont
  {Boccardo}, \citenamefont {Benamara},\ and\ \citenamefont
  {Pierre-Louis}}]{Boccardo2022b}%
  \BibitemOpen
  \bibfield  {author} {\bibinfo {author} {\bibfnamefont {F.}~\bibnamefont
  {Boccardo}}, \bibinfo {author} {\bibfnamefont {Y.}~\bibnamefont {Benamara}},\
  and\ \bibinfo {author} {\bibfnamefont {O.}~\bibnamefont {Pierre-Louis}},\
  }\href {https://doi.org/10.1103/PhysRevE.106.024120} {\bibfield  {journal}
  {\bibinfo  {journal} {Phys. Rev. E}\ }\textbf {\bibinfo {volume} {106}},\
  \bibinfo {pages} {024120} (\bibinfo {year} {2022})}\BibitemShut {NoStop}%
\bibitem [{\citenamefont {Swope}\ \emph {et~al.}(2004)\citenamefont {Swope},
  \citenamefont {Pitera},\ and\ \citenamefont {Suits}}]{Swope2004}%
  \BibitemOpen
  \bibfield  {author} {\bibinfo {author} {\bibfnamefont {W.~C.}\ \bibnamefont
  {Swope}}, \bibinfo {author} {\bibfnamefont {J.~W.}\ \bibnamefont {Pitera}},\
  and\ \bibinfo {author} {\bibfnamefont {F.}~\bibnamefont {Suits}},\
  }\href@noop {} {\bibfield  {journal} {\bibinfo  {journal} {The Journal of
  Physical Chemistry B}\ }\textbf {\bibinfo {volume} {108}},\ \bibinfo {pages}
  {6571} (\bibinfo {year} {2004})}\BibitemShut {NoStop}%
\bibitem [{\citenamefont {Bellman}(2003)}]{Bellman1957}%
  \BibitemOpen
  \bibfield  {author} {\bibinfo {author} {\bibfnamefont {R.~E.}\ \bibnamefont
  {Bellman}},\ }\href@noop {} {\emph {\bibinfo {title} {Dynamic Programming}}}\
  (\bibinfo  {publisher} {Dover Publications, Inc.},\ \bibinfo {address}
  {USA},\ \bibinfo {year} {2003})\BibitemShut {NoStop}%
\bibitem [{\citenamefont {Boccardo}\ and\ \citenamefont
  {Pierre-Louis}(2022{\natexlab{b}})}]{Boccardo2022c}%
  \BibitemOpen
  \bibfield  {author} {\bibinfo {author} {\bibfnamefont {F.}~\bibnamefont
  {Boccardo}}\ and\ \bibinfo {author} {\bibfnamefont {O.}~\bibnamefont
  {Pierre-Louis}},\ }\href {https://doi.org/10.1088/1742-5468/ac9616}
  {\bibfield  {journal} {\bibinfo  {journal} {Journal of Statistical Mechanics:
  Theory and Experiment}\ }\textbf {\bibinfo {volume} {2022}},\ \bibinfo
  {pages} {103205} (\bibinfo {year} {2022}{\natexlab{b}})}\BibitemShut
  {NoStop}%
\bibitem [{\citenamefont {Baronchelli}\ and\ \citenamefont
  {Loreto}(2006)}]{Baronchelli2006}%
  \BibitemOpen
  \bibfield  {author} {\bibinfo {author} {\bibfnamefont {A.}~\bibnamefont
  {Baronchelli}}\ and\ \bibinfo {author} {\bibfnamefont {V.}~\bibnamefont
  {Loreto}},\ }\href {https://doi.org/10.1103/PhysRevE.73.026103} {\bibfield
  {journal} {\bibinfo  {journal} {Phys. Rev. E}\ }\textbf {\bibinfo {volume}
  {73}},\ \bibinfo {pages} {026103} (\bibinfo {year} {2006})}\BibitemShut
  {NoStop}%
\bibitem [{\citenamefont {Khare}\ \emph {et~al.}(1995)\citenamefont {Khare},
  \citenamefont {Bartelt},\ and\ \citenamefont {Einstein}}]{Khare1995}%
  \BibitemOpen
  \bibfield  {author} {\bibinfo {author} {\bibfnamefont {S.~V.}\ \bibnamefont
  {Khare}}, \bibinfo {author} {\bibfnamefont {N.~C.}\ \bibnamefont {Bartelt}},\
  and\ \bibinfo {author} {\bibfnamefont {T.~L.}\ \bibnamefont {Einstein}},\
  }\href {https://doi.org/10.1103/PhysRevLett.75.2148} {\bibfield  {journal}
  {\bibinfo  {journal} {Phys. Rev. Lett.}\ }\textbf {\bibinfo {volume} {75}},\
  \bibinfo {pages} {2148} (\bibinfo {year} {1995})}\BibitemShut {NoStop}%
\bibitem [{\citenamefont {Pierre-Louis}(2001)}]{PierreLouis2001}%
  \BibitemOpen
  \bibfield  {author} {\bibinfo {author} {\bibfnamefont {O.}~\bibnamefont
  {Pierre-Louis}},\ }\href {https://doi.org/10.1103/PhysRevLett.87.106104}
  {\bibfield  {journal} {\bibinfo  {journal} {Phys. Rev. Lett.}\ }\textbf
  {\bibinfo {volume} {87}},\ \bibinfo {pages} {106104} (\bibinfo {year}
  {2001})}\BibitemShut {NoStop}%
\bibitem [{\citenamefont {Giesen-Seibert}\ \emph {et~al.}(1995)\citenamefont
  {Giesen-Seibert}, \citenamefont {Schmitz}, \citenamefont {Jentjens},\ and\
  \citenamefont {Ibach}}]{Giesen1995}%
  \BibitemOpen
  \bibfield  {author} {\bibinfo {author} {\bibfnamefont {M.}~\bibnamefont
  {Giesen-Seibert}}, \bibinfo {author} {\bibfnamefont {F.}~\bibnamefont
  {Schmitz}}, \bibinfo {author} {\bibfnamefont {R.}~\bibnamefont {Jentjens}},\
  and\ \bibinfo {author} {\bibfnamefont {H.}~\bibnamefont {Ibach}},\ }\href
  {https://doi.org/https://doi.org/10.1016/0039-6028(95)00055-0} {\bibfield
  {journal} {\bibinfo  {journal} {Surface Science}\ }\textbf {\bibinfo {volume}
  {329}},\ \bibinfo {pages} {47} (\bibinfo {year} {1995})}\BibitemShut
  {NoStop}%
\bibitem [{\citenamefont {Igl}\ \emph {et~al.}(2019)\citenamefont {Igl},
  \citenamefont {Ciosek}, \citenamefont {Li}, \citenamefont {Tschiatschek},
  \citenamefont {Zhang}, \citenamefont {Devlin},\ and\ \citenamefont
  {Hofmann}}]{Igl2019}%
  \BibitemOpen
  \bibfield  {author} {\bibinfo {author} {\bibfnamefont {M.}~\bibnamefont
  {Igl}}, \bibinfo {author} {\bibfnamefont {K.}~\bibnamefont {Ciosek}},
  \bibinfo {author} {\bibfnamefont {Y.}~\bibnamefont {Li}}, \bibinfo {author}
  {\bibfnamefont {S.}~\bibnamefont {Tschiatschek}}, \bibinfo {author}
  {\bibfnamefont {C.}~\bibnamefont {Zhang}}, \bibinfo {author} {\bibfnamefont
  {S.}~\bibnamefont {Devlin}},\ and\ \bibinfo {author} {\bibfnamefont
  {K.}~\bibnamefont {Hofmann}},\ }\href@noop {} {\bibfield  {journal} {\bibinfo
   {journal} {Advances in neural information processing systems}\ }\textbf
  {\bibinfo {volume} {32}} (\bibinfo {year} {2019})}\BibitemShut {NoStop}%
\bibitem [{\citenamefont {Arjovsky}\ and\ \citenamefont
  {Bottou}(2017)}]{Arjovsky2017}%
  \BibitemOpen
  \bibfield  {author} {\bibinfo {author} {\bibfnamefont {M.}~\bibnamefont
  {Arjovsky}}\ and\ \bibinfo {author} {\bibfnamefont {L.}~\bibnamefont
  {Bottou}},\ }\href@noop {} {\bibinfo {title} {Towards principled methods for
  training generative adversarial networks}} (\bibinfo {year} {2017}),\ \Eprint
  {https://arxiv.org/abs/1701.04862} {arXiv:1701.04862 [stat.ML]} \BibitemShut
  {NoStop}%
\bibitem [{\citenamefont {Barfuss}\ and\ \citenamefont
  {Meylahn}(2023)}]{Barfuss2023}%
  \BibitemOpen
  \bibfield  {author} {\bibinfo {author} {\bibfnamefont {W.}~\bibnamefont
  {Barfuss}}\ and\ \bibinfo {author} {\bibfnamefont {J.~M.}\ \bibnamefont
  {Meylahn}},\ }\href@noop {} {\bibfield  {journal} {\bibinfo  {journal}
  {Scientific Reports}\ }\textbf {\bibinfo {volume} {13}},\ \bibinfo {pages}
  {1309} (\bibinfo {year} {2023})}\BibitemShut {NoStop}%
\bibitem [{\citenamefont {van Hasselt}\ \emph {et~al.}(2016)\citenamefont {van
  Hasselt}, \citenamefont {Guez},\ and\ \citenamefont
  {Silver}}]{VanHasselt2016}%
  \BibitemOpen
  \bibfield  {author} {\bibinfo {author} {\bibfnamefont {H.}~\bibnamefont {van
  Hasselt}}, \bibinfo {author} {\bibfnamefont {A.}~\bibnamefont {Guez}},\ and\
  \bibinfo {author} {\bibfnamefont {D.}~\bibnamefont {Silver}},\ }\href
  {https://doi.org/10.1609/aaai.v30i1.10295} {\bibfield  {journal} {\bibinfo
  {journal} {Proceedings of the AAAI Conference on Artificial Intelligence}\
  }\textbf {\bibinfo {volume} {30}},\  (\bibinfo {year} {2016})}\BibitemShut
  {NoStop}%
\bibitem [{\citenamefont {Taylor}\ and\ \citenamefont
  {Stone}(2009)}]{Taylor2009}%
  \BibitemOpen
  \bibfield  {author} {\bibinfo {author} {\bibfnamefont {M.~E.}\ \bibnamefont
  {Taylor}}\ and\ \bibinfo {author} {\bibfnamefont {P.}~\bibnamefont {Stone}},\
  }\href {https://www.jmlr.org/papers/volume10/taylor09a/taylor09a.pdf}
  {\bibfield  {journal} {\bibinfo  {journal} {Journal of Machine Learning
  Research}\ }\textbf {\bibinfo {volume} {10}},\  (\bibinfo {year}
  {2009})}\BibitemShut {NoStop}%
\bibitem [{\citenamefont {Zhuang}\ \emph {et~al.}(2020)\citenamefont {Zhuang},
  \citenamefont {Qi}, \citenamefont {Duan}, \citenamefont {Xi}, \citenamefont
  {Zhu}, \citenamefont {Zhu}, \citenamefont {Xiong},\ and\ \citenamefont
  {He}}]{Zhuang2020}%
  \BibitemOpen
  \bibfield  {author} {\bibinfo {author} {\bibfnamefont {F.}~\bibnamefont
  {Zhuang}}, \bibinfo {author} {\bibfnamefont {Z.}~\bibnamefont {Qi}}, \bibinfo
  {author} {\bibfnamefont {K.}~\bibnamefont {Duan}}, \bibinfo {author}
  {\bibfnamefont {D.}~\bibnamefont {Xi}}, \bibinfo {author} {\bibfnamefont
  {Y.}~\bibnamefont {Zhu}}, \bibinfo {author} {\bibfnamefont {H.}~\bibnamefont
  {Zhu}}, \bibinfo {author} {\bibfnamefont {H.}~\bibnamefont {Xiong}},\ and\
  \bibinfo {author} {\bibfnamefont {g.}~\bibnamefont {He}, \bibfnamefont
  {Qin}},\ }\href {https://doi.org/10.1109/jproc.2020.3004555} {\bibfield
  {journal} {\bibinfo  {journal} {Proceedings of the IEEE}\ }\textbf {\bibinfo
  {volume} {109}},\ \bibinfo {pages} {43} (\bibinfo {year} {2020})}\BibitemShut
  {NoStop}%
\end{thebibliography}%

\end{document}